\documentclass{ifacconf}
\usepackage{amsmath,amsfonts,amssymb}
\usepackage{mathtools}
\usepackage{bm}
\usepackage{color,xcolor}
\usepackage{graphicx,tikz}      
\usepackage{natbib}        
\usepackage{url,tabularx}
\newcommand{\argmin}[1]{\underset{#1}{\operatorname{min}}}

\makeatletter
\newcommand{\bigcircledast}{\mathop{\mathpalette\b@gcircledast\relax}}
\newcommand{\b@gcircledast}[2]{%
  \vcenter{\hbox{\m@th
    \scalebox{\ifx#1\displaystyle 2.1\else1.4\fi}{$#2\circledast$}%
  }}%
}
\makeatother

\begin{document}
\begin{frontmatter}

\title{Tensor Network Kernel Machines for Nonlinear System Identification\\[-6mm]}


\author[First]{Kim Batselier}

\address[First]{Delft University of Technology, Delft, The Netherlands \\ (e-mail: k.batselier@tudelft.nl).}


\begin{abstract}                
Over the past ten years Tensor Network Kernel Machines have enabled efficient nonlinear system identification through the use of low-rank tensor networks. The goal of this article is to provide the reader with a basic understanding of this model class in terms of its two defining properties. Furthermore, both the optimization and regularization of such models are discussed, together with recent developments and starting points for further exploration.
\end{abstract}


\end{frontmatter}

\section{Introduction}
Kernel machines have a rich history of being applied to system identification problems\footnote{This work has been submitted to IFAC for possible publication.}. The first occurrence of their use was in 1998 with Support Vector Machines (SVM)~\citep{drezet1998support}. The reformulation of SVMs into Least Squares-Support Vector Machines (LS-SVM)~\citep{suykens2002least} was a seminal result in the sense that it replaced a Quadratic Program with a simple linear least squares problem. Consequently, LS-SVMs quickly found their use in system identification. Among the applications are partially linear systems~\citep{espinoza2005kernel}, Hammerstein systems~\citep{goethals2005subspace}, Wiener-Hammerstein systems~\citep{falck2009identification}, and LPV-ARX models~\citep{toth2011model}, to name a few.

At around the same time when LS-SVMs were being developed, Gaussian Processes (GPs) attracted a lot of attention in the machine learning community~\citep{williams2006gaussian}. The main difference between GPs and LS-SVMs comes from the former being derived from a probabilistic point of view. The main advantage of GPs is that confidence bounds on predictions become straightforward to compute. Just like their deterministic LS-SVM counterparts, GPs have found ample use in system identification~\citep{sarkka2021use}.

The main advantage of kernel machines is their primal-dual formulation, giving rise to a great amount of flexibility. The primal formulation considers a feature map $\varphi(\bm{x})$ that maps $\bm{x} \in \mathbb{R}^{D}$ to some $P$-dimensional  (possibly infinite) feature space. The dual formulation on the other hand considers a kernel function $\kappa(\bm{x}_i,\bm{x}_j)$ that computes some measure of similarity between data points $\bm{x}_i$ and $\bm{x}_j$. The zero duality gap then implies that the solutions of the primal and dual problem are identical, thus giving rise to the aforementioned flexibility. When the sample size $N$ is much larger than the dimension of the feature space $P$, then solving the primal problem is computationally advantageous with a complexity $O(P^3)$. When $P$ is much larger than $N$, then solving the dual problem is more advantageous with a complexity $O(N^3)$.

But what if both $N$ and $P$ are so large that neither the primal nor the dual problem can be solved? A Tensor Network Kernel Machine (TNKM) solves exactly this problem~\citep{batselier_2022}. The key ingredient for TNKM's success lies in the combination of a tensor network (TN) parameterization with a Kronecker product structured feature vector. By exploiting these structures the nonlinear system identification problem can be solved for \mbox{$P=I^D$} parameters and $N$ samples with a computational complexity of $O(NDIR + I^3R^3)$, rather than $O(I^{3D})$ or $O(N^3)$.

The main content of this article is divided into three sections. In Section~\ref{sec:model} the TNKM model is discussed in terms of its two defining characteristics: a Kronecker product structure combined with a TN parameterization. Section~\ref{sec:opt} goes into the optimization and regularization of TNKMs. Finally, Section~\ref{sec:dev} mentions recent TNKM developments.

\section{Tensor network kernel machines}
\label{sec:model}
\subsection{Notation}
A tensor $\bm{\mathcal{A}} \in \mathbb{R}^{I_1 \times I_2 \times \cdots \times I_D}$ is a multidimensional array, always denoted by a calligraphic uppercase boldface letter. Scalars are denoted by lowercase letters, vectors by lowercase boldface and matrices by uppercase boldface. The $D$ entries of a vector $\bm{x} \in \mathbb{R}^D$ are written as $x_1,\ldots,x_D$. One-based indexing is used and  the maximal value an index can attain is always its corresponding uppercase letter. For example, the index $i$ satisfies $1 \leq i \leq I$.

\subsection{The Model}
To keep notation simple we only consider the single output case throughout this article. Consider a nonlinear system identification problem with data $\bm{\mathcal{D}} \coloneq \{(\bm{x}_n,y_n) \}_{n=1}^N$. The vector $\bm{x}_n$ contains all past information up to time \mbox{$n-1$}, necessary to predict the output $y_n$ at time $n$. Before defining a TNKM, we first consider the SVM/GP regression model
\begin{align}
\label{eq:model}
    y_n &= \hat{y}(\bm{x}_n) + e_n = \bm{\varphi}(\bm{x}_n)^\intercal\,\bm{w} + e_n.
\end{align}
The nonlinear 1-step ahead predictor $\hat{y}(\bm{x}_n)$ is parameterized as the dot product between a feature vector $\bm{\varphi}(\bm{x}_n)$ and model parameters $\bm{w}$. The white noise disturbance process is denoted $e_n$. 

A TNKM is defined by model equation~\eqref{eq:model} and the following two properties:
\begin{enumerate}
    \item $\bm{\varphi}(\bm{x}_n) \in \mathbb{R}^{I^D}$ has a Kronecker product structure
    \begin{align}
    \label{eq:kronphi}
    \bm{\varphi}(\bm{x}_n) \coloneq \bm{\varphi}_D \otimes \cdots \otimes\bm{\varphi}_2 \otimes  \bm{\varphi}_{1}.
    \end{align}
    \item $\bm{w}  \in \mathbb{R}^{I^D}$ is parameterized as a tensor network.
\end{enumerate}
These two properties complement one another, enabling TNKMs to identify highly-nonlinear dynamical systems very efficiently. The reason for this efficiency is that instead of solving the identification problem for $\bm{w}$, only its tensor network components need to be identified instead. We unpack these two defining properties in the next two subsections.

\subsection{Kronecker product feature vector $\bm{\varphi({x}}_n)$}
The feature vector $\bm{\varphi}(\bm{x}_n)$ contains a collection of multivariate basis functions in the variables $\bm{x}_n$. The nonlinear 1-step ahead predictor $\hat{y}(\bm{x}_n)$ is then obtained through a linear combination of these basis functions with appropriate weights $\bm{w}$. The Kronecker product structure in~\eqref{eq:kronphi} is merely a way to construct $I^D$ multivariate basis functions from products of $I$ univariate basis functions.

Different choices of $\bm{x}_n$ result in different model structures. A direct feedthrough term can be easily added through the inclusion of $u_n$ into $\bm{x}$. An overview of common choices together with the resulting model structures are given in Table~\ref{tab:choices}. All of these features can easily describe multi-input systems by replacing all scalars $u_{n-1}, u_{n-2}$ in $\bm{x}$ with their corresponding vectors. TNKMs are not restricted to input-output parameterizations. Nonlinear state space models of the form $ \dot{\bm{x}} = f(\bm{x})$, with fully observed state information are TNKMs when choosing $y=\dot{\bm{x}}$. A well-developed model in this nonlinear state space setting that uses this idea is the multilinear time-invariant (MTI) model~\citep{Lichtenberg2011Hybrid}.

We now discuss some common choices for basis functions. For notational clarity we drop the discrete-time subscript notation such that $\bm{x}_n$ is replaced with $\bm{x}$ from now on. 
\begin{table}[]
    \centering
    \begin{tabular}{c c}
        $\bm{x}_n$ & Model Structure\\
        \hline
        $u_{n-1},u_{n-2},\ldots$ &  NFIR\\
        $y_{n-1},y_{n-2},\ldots$ &  NAR\\        
        $u_{n-1},u_{n-2},\ldots,y_{n-1},y_{n-2},\ldots$ &  NARX\\
        $u_{n-1},u_{n-2},\ldots,\hat{y}_{n-1},\hat{y}_{n-2},\ldots$ &  NOE\\        
        $u_{n-1},u_{n-2},\ldots,y_{n-1},y_{n-2},\ldots,e_{n-1},e_{n-2},\ldots$ &  NARMAX\\
        &
    \end{tabular}
    \caption{Common choices for $\bm{x}_n$ in a TNKM, resulting in different nonlinear model structures.}
    \label{tab:choices}
\end{table}

\textbf{Monomial basis functions.} 
The Stone-Weierstrass theorem \citep{de_branges_stone-weierstrass_1959} guarantees that any continuous function on a locally compact domain can be approximated arbitrarily well by polynomials of increasing degree. This fact serves as a motivation to approximate the 1-step ahead predictor $\hat{y}(\bm{x})$ by a multivariate polynomial of total degree $D$. All monomials from total degree 0 up to $D$ can be constructed through the Kronecker product~\citep{batselier2017tensor}
\begin{align}
\label{def:mon}
\bm{\varphi}(\bm{x})^\intercal\coloneq \underbrace{
    \begin{pmatrix}
        1 & \bm{x}
    \end{pmatrix} \otimes \cdots \otimes \begin{pmatrix}
        1 & \bm{x}
    \end{pmatrix}}_{D \textrm{ times}}.
\end{align}
Setting $\bm{x}$ equal to $M$ past inputs $u_{n-1},\ldots,u_{n-M}$ results in a truncated Volterra model~\citep{batselier2017tensor}. The monomial basis feature vector in~\eqref{def:mon} is equivalent with the polynomial kernel since $    \bm{\varphi}(\bm{x})^\intercal \,\bm{\varphi}(\bm{x}) = (1+\bm{x}^\intercal\bm{x})^D$. In the context of TNKMs, however, we will not make further use of kernel functions since the whole idea will be to impose a tensor network structure onto $\bm{w}$ and exploit this
structure for storage and computational gains.

\textbf{Pure-power polynomial basis functions.} Pure-power polynomial features~\citep{chen2017parallelized} are an alternative to construct polynomial nonlinearities. For a vector \mbox{$\bm{x} \in \mathbb{R}^D$} they are defined as
\begin{align*}
    \bm{\varphi}(\bm{x})^\intercal &\coloneq \bigotimes_{d=1}^D \begin{pmatrix}
        1 & x_d & x_d^2 & \cdots & x_d^{I-1} 
    \end{pmatrix}, 
\end{align*}
where $x_d$ denotes the $d$th entry of $\bm{x}$. Each factor in the Kronecker product therefore consists of a specific $x_d$, in contrast to the monomial basis where all $x_d$ components are present in each factor. Generating  $I^D$ monomials using pure-power polynomial features results in more higher-order terms compared to the monomial basis functions. Similarly, the Stone-Weierstrass theorem applies. 

\textbf{Fourier Features.} Two kinds of Kronecker product Fourier Features have been described in the literature, both approximate the Gaussian kernel function. The first option~\citep{wesel2021large} considers the domain of the 1-step ahead predictor to be the hypercube $[-U,U]^D$ with vanishing boundary conditions. The $i$th entry of the $d$th feature vector in that case is defined as
\begin{align*}
    (\bm{\varphi}_d)_i &\coloneq \frac{1}{\sqrt{U}}  \; p\left(\frac{i\,\pi}{2U}\right) \; \sin{\left(\frac{i\,\pi(x_d+U)}{2U} \right)}, 
\end{align*}    
where $p(\cdot)$ is the spectral density of the one-dimensional Gaussian kernel. Each $\bm{\varphi}_d$ vector has $I$ associated frequencies which are mixed via products into a total of $I^D$ frequencies. Removing the zero boundary conditions on the domain requires adding $I$ cosine contributions. Alternatively, an 1-step ahead predictor with period $L \in \mathbb{R}$~\citep{wesel2024quantized} can be modeled through the features
\begin{align*}
(\bm{\varphi}_d)_i &\coloneq e^{2\pi \, j\,x_d \, \frac{2+I}{2L}}\; \left(e^{-\frac{2\pi \, j \, x_d}{L}}\right)^{(i-1)},     
\end{align*}
with $j$ the imaginary unit and again $I$ frequencies associated with each feature vector $\bm{\varphi}_d$.

\textbf{Quantized features.} The pure-power polynomial and complex exponential Fourier features can be represented even more efficiently~\citep{wesel2024quantized} when the total number of basis functions $I$ satisfies $I=Q^K$, with $Q,K$ positive integers. The features in this case can be further decomposed into a Kronecker product, giving rise to so-called quantized features. For example, when $I=2^3$ then
\begin{align*}
    \begin{pmatrix}
        1 & x_d & x_d^2 & \cdots & x_d^7
    \end{pmatrix} = \begin{pmatrix}1 & x_d \end{pmatrix} \otimes \begin{pmatrix}1 & x_d^2 \end{pmatrix} \otimes \begin{pmatrix}1 & x_d^4 \end{pmatrix}.
\end{align*}
A similar relation holds for the complex exponential Fourier features. This result has major consequences for the parameterization of $\bm{w}$. A theoretical derivation through the VC-dimension and various numerical experiments show that the parameterization of $\bm{w}$ that uses quantized features have better generalization properties compared to their conventional counterparts~\citep{wesel2024quantized}.

\textbf{B-splines.} B-splines are piecewise polynomials that can approximate the 1-step ahead predictor in a hypercube $[-U,U]^D$ up to any desired accuracy. Each dimension of the hypercube is first uniformly divided into $M$ intervals $t_m \leq x \leq t_{m+1}$, also called the knot sequence $\{t_0, t_1, \ldots, t_{M-1}, t_M\}$. This division determines the accuracy of the approximation, where larger $M$ results in a better approximation. A B-spline basis function $B_{i,\rho}$ is a polynomial of degree $\rho$ that remains nonzero over \mbox{$\rho+1$} consecutive intervals. The $d$th feature vector is hence defined as
\begin{align*}
    (\bm{\varphi}_d)^\intercal &\coloneq   \begin{pmatrix}B_{1,\rho}(x_d) & B_{2,\rho}(x_d) & \cdots & B_{I,\rho}(x_d) \end{pmatrix}, 
\end{align*}  
with a total of $I=M-\rho$ basis functions. A nice property of these basis functions is that the smoothness of the resulting 1-step ahead predictor can be directly controlled by limiting the difference between consecutive weights through regularization~\citep{karagoz2020nonlinear}. Furthermore, such regularization can be encoded as a low-rank tensor network, which enables an efficient implementation.

\subsection{Tensor Network Parameterization of $\bm{w}$}
The second defining property of a TNKM is that $\bm{w}  \in \mathbb{R}^{I^D}$ is parameterized by a tensor network. For the context of this article, a tensor network $\textrm{TN}(\cdot,\ldots,\cdot)$ is a multilinear function of $D$ TN-cores $\bm{\mathcal{W}}_1,\ldots,\bm{\mathcal{W}}_D$ such that
\begin{align}
\label{def:TN}
    \bm{w} = \textrm{TN}(\bm{\mathcal{W}}_1,\bm{\mathcal{W}}_2,\ldots,\bm{\mathcal{W}}_D).
\end{align}
TN-cores are small tensors that can be interpreted as fully describing $\bm{w}$.    A function is multilinear when it is a linear function in any $\bm{\mathcal{W}}_d$ whilst keeping all other cores fixed. Therefore, equation~\eqref{def:TN} can always be rewritten as
\begin{align}
\label{def2:TN}
    \bm{w} &= \bm{U}_d \; \textrm{vec}(\bm{\mathcal{W}}_d),
\end{align}
for any of the $D$ TN-cores $\bm{\mathcal{W}}_d$. The matrix $\bm{U}_d$ depends on all other TN-cores $k \neq d$, and the $\textrm{vec}(\cdot)$ operation stacks all entries of $\bm{\mathcal{W}}_d$ into a vector. The storage complexity of a TN is simply the storage complexity of the $D$ TN-cores. For example, when each TN-core is a matrix of size $I \times R$, then the total storage cost of the TN is $DIR$. For low $R$, this constitutes a massive saving compared to the explicit storage of the $I^D$ entries of $\bm{w}$.

By replacing $\bm{w}$ in $\hat{y}_n$ with a TN, the $I^D$ entries in $\bm{w}$ are never explicitly stored and the estimation of $\bm{w}$ becomes the estimation of the TN-cores $\bm{\mathcal{W}}_d$ instead. In addition to obvious storage gains, significant computational gains can also be obtained through the development of optimization algorithms that exploit the tensor network structure of $\hat{y}_n$.
We now discuss the two most commonly-used tensor networks in TNKMs.

\textbf{Canonical Polyadic Decomposition (CPD).} The TN-cores of a CPD are matrices $\bm{W}_d \in \mathbb{R}^{I \times R}$
such that
\begin{align}
\label{def:CPD}
\bm{w} &=  \sum_{r=1}^R \bm{w}_D^r \otimes \cdots \otimes \bm{w}_1^r,
\end{align}
where $\bm{w}_d^r$ denotes the vector obtained from setting the index $r$ in $\bm{W}_d$ to a fixed value, and $R$ is called the CPD-rank~\citep{Hi27}. The CPD-rank directly determines the ``richness" of the set of $\bm{w}$ vectors that are described by the CPD. Increasing $R$ always decreases the loss on identification data but at some point overfitting will occur. In Section~\ref{sec:opt} we will explain how overfitting can be countered through regularization. 

One of the main benefits of using a CPD now becomes clear, the original $I^D$  storage cost is reduced to $DIR$, since only $D$ matrices $\bm{W}_d$, each of size $I \times R$, need to be stored. The equivalent of equation~\eqref{def2:TN} for the CPD is
\begin{align}
\label{def:CPD2}
    \bm{w} &= \underbrace{\left( \bigodot_{k\neq d} \bm{W}_k  \otimes \bm{I}_I\right)}_{\bm{U}_d} \; \textrm{vec}(\bm{W}_d),
\end{align}
with $\odot$ the column-wise Kronecker product and $\bm{I}_I$ the $I \times I$ identity matrix. Substituting equation~\eqref{def:CPD2} into equation~\eqref{eq:model} together with the Kronecker product for $\bm{\varphi}$ results in an expression for the 1-step ahead predictor that is linear in $\bm{W}_d$
\begin{align*}
    \hat{y}_n &=\left(\bm{\varphi}_D \otimes \cdots \otimes \bm{\varphi}_1 \right)^\intercal \;\left( \bigodot_{k\neq d} \bm{W}_k  \otimes \bm{I}_I\right)\; \textrm{vec}(\bm{W}_d),\\
    &= \left( \bigcircledast_{k\neq d}\; \bm{\varphi}_k^\intercal\;\bm{W}_k  \otimes \bm{\varphi}_d^\intercal \right)\; \textrm{vec}(\bm{W}_d),
\end{align*}
where $\bigcircledast$ is the elementwise (Hadamard) product. The computational cost for evaluating the conventional predictor $\hat{y}_n$ for $N$ samples is $O(NI^D)$, while for the TNKM it is $O(NDIR)$~\citep{Lichtenberg2011Hybrid,wesel2021large}. The exponential complexity has been completely resolved! 

\textbf{Tensor Train (TT).} The TN-cores of a TT are 3rd-order tensors $\bm{\mathcal{W}}_d \in \mathbb{R}^{R_{d-1} \times I \times R_{d}}$ such that
\begin{align}
    \label{def:TT}
    \bm{w} &= \sum_{r_1=1}^{R_1} \cdots \sum_{r_{D-1}=1}^{R_{D-1}} \bm{w}_D^{r_{D-1},r_D} \otimes \cdots \otimes  \bm{w}_1^{r_{0},r_1},
\end{align}
where $\bm{w}_d^{r_{d-1},r_d}$ denotes the vector obtained from setting the $r_{d-1}$ and $r_d$ indices in $\bm{\mathcal{W}}_d$ to fixed values~\citep{Os11}. In addition, boundary conditions $R_0=R_D=1$ are required for the right-hand side in equation~\eqref{def:TT} to be a vector. The most notable difference with the CPD is that a TT has $D-1$ different TT-ranks $R_1,R_2,\ldots,R_{D-1}$. The total storage cost of a TT is $O(DIR^2)$, assuming equal TT-ranks. This multitude of ranks gives the TT more flexibility in describing $\bm{w}$, since a total of $DIR$ TN-parameters can be attained by possibly several different configurations of TT-ranks. In contrast, only a CPD-rank $R$ results in a total number of $DIR$ TN-parameters. The same argument on overfitting applies to TTs.

In order to derive equation~\eqref{def2:TN} for TTs, we first need the following intermediate definitions
\begin{align*}
    \bm{W}_{<d} &\coloneq \sum_{r_1=1}^{R_1} \cdots \sum_{r_{d-2}=1}^{R_{d-2}} \bm{w}_{d-1}^{r_{d-2},r_{d-1}} \otimes \cdots \otimes  \bm{w}_1^{r_{0},r_1},\\
    \bm{W}_{>d} &\coloneq \sum_{r_{d+1}=1}^{R_{d+1}} \cdots \sum_{r_{D-1}=1}^{R_{D-1}} \bm{w}_D^{r_{D-1},r_D} \otimes \cdots \otimes  \bm{w}_{d+1}^{r_{d},r_{d+1}},
\end{align*}
From these definitions it follows that $\bm{W}_{<d} \in \mathbb{R}^{I^{d-1}\times R_{d-1}}$ and $\bm{W}_{>d} \in \mathbb{R}^{I^{D-d}\times R_{d}}$. Together with the boundary conditions $\bm{W}_{<1}=\bm{W}_{>d}=1$ we can then write
\begin{align}
    \label{def:TT2}
 \bm{w} = \left(\bm{W}_{>d} \otimes \bm{I}_{I} \otimes \bm{W}_{<d} \right) \; \textrm{vec}(\bm{\mathcal{W}}_d)   
\end{align}
In order to derive the 1-step ahead predictor $\hat{y}_n$ using a TT parameterization of $\bm{w}$ we first need similar intermediate definitions for the feature vector
\begin{align*}
    \bm{\varphi}_{<d} &\coloneq  \bm{\varphi}_{d-1} \otimes \cdots \otimes  \bm{\varphi}_1 \in \mathbb{R}^{I^{d-1}},\\
    \bm{\varphi}_{>d} &\coloneq \bm{\varphi}_D \otimes \cdots \otimes  \bm{\varphi}_{d+1} \in \mathbb{R}^{I^{D-d}}.
\end{align*}
These allow us to write
\begin{align*}
    \hat{y}_n &= \left(\bm{\varphi}_{>d}\otimes \bm{\varphi}_d \otimes \bm{\varphi}_{<d} \right)^\intercal \;\left(\bm{W}_{>d} \otimes \bm{I}_{I} \otimes \bm{W}_{<d} \right) \; \textrm{vec}(\bm{\mathcal{W}}_d) ,\\
    &= \left(\bm{\varphi}_{>d}^\intercal\bm{W}_{>d} \otimes \bm{\varphi}_d^\intercal \otimes \bm{\varphi}_{<d}^\intercal\bm{W}_{<d} \right)\; \textrm{vec}(\bm{\mathcal{W}}_d).
\end{align*}   

The TT parameterization of $\bm{w}$ allows the 1-step ahead predictor to be evaluated for $N$ samples with a computational complexity $O(NDIR^2)$~\citep{batselier2017tensor}.

\section{Optimization and Regularization}
\label{sec:opt}
Two choices are essential when identifying nonlinear systems as TNKMs. The first choice is the model structure, which is completely determined through $\bm{x}$. The second choice is the TN parameterization of $\bm{w}$. An estimate for $\bm{w}$ is found from solving
\begin{align}
    \label{def:opti}
   \argmin{\bm{\mathcal{W}}_1,\ldots,\bm{\mathcal{W}}_D}&\; \frac{1}{N} \sum_{n=1}^N \left\lVert y_n - \hat{y}_n \right\rVert_2^2 + \mathcal{R}(\bm{\mathcal{W}}_1,\ldots,\bm{\mathcal{W}}_D),
\end{align}
where both the model fit term and regularization term depend on the TN-cores $\bm{\mathcal{W}}_1,\ldots,\bm{\mathcal{W}}_D$. The $\ell_2$-norm for the model fit term is the most common choice, but other choices are possible. In contrast to the conventional LS-SVM or GP model, the TNKM 1-step ahead predictor $\hat{y}_n$ is a multilinear function of the TN-cores, rendering the optimization problem nonlinear.

\subsection{Optimization}
In principle, one can use any off-the-shelf method for solving~\eqref{def:opti}. Common choices when $N$ is large are Stochastic Gradient Descent~\citep{bottou2018optimization} and the Adam optimizer~\citep{kingma2014adam}. These methods come with great flexibility and ease-of-use, but at the cost of not being specifically tailored to exploit the existing tensor network structure. Updating all TN-cores at once also becomes computationally expensive for larger $D$ and TN-ranks $R$.

Exploiting this particular multilinear structure can be easily done with a nonlinear Gauss Seidel method, also called block coordinate descent. The main idea is to initialize the TN-cores and update only one keeping all others fixed. The multilinear structure turns this update into a simple linear least squares problem, which can be repeated for all TN-cores until convergence. The name for this procedure is ALS, which stands for alternating least squares or alternating linear scheme~\citep{holtz2012alternating,uschmajew2012local}.

Derivation of the linear system for a TN-core update due to the model fit term in~\eqref{def:opti} is straightforward. Let $\bm{\Phi} \in \mathbb{R}^{N \times I^D}$ be the matrix where row $n$ is the feature vector $\bm{\varphi}(\bm{x}_n)^\intercal$, and $\bm{y} \in \mathbb{R}^N$ the vector of outputs. The model fit term is then\;
\begin{align*}
\frac{1}{N}    \left(\bm{y}-\bm{\Phi}\bm{U}_d\;\textrm{vec}(\bm{\mathcal{W}}_d)\right)^\intercal\;\left((\bm{y}-\bm{\Phi}\bm{U}_d\;\textrm{vec}(\bm{\mathcal{W}}_d)\right),
\end{align*}
leading to the normal equations
\begin{align*}
    \bm{U}_d^\intercal\bm{\Phi}^\intercal\bm{\Phi}\bm{U}_d\;\textrm{vec}(\bm{\mathcal{W}}_d) &= \bm{U}_d^\intercal\bm{\Phi}^\intercal\bm{y},
\end{align*}
which can be solved analytically in $O(I^3R^3)$ flops for the CPD and $O(I^3R^6)$ flops for the TT.

Combining this ALS procedure with gradient boosting results in a fast automatic determination of the TN-ranks removing the need for cross-validation, together with reduced peak memory usage~\citep{saiapin2026boosting}. The regularization term $\mathcal{R}$ adds an additional term to the left-hand side of the normal equations but does not affect the computational complexity in terms of $O(\cdot)$ flop count. We now discuss some commonly-used choices.

\subsection{Regularization}
Several different choices for the regularization term $\mathcal{R}$ have been described in the literature. From a Bayesian perspective, each of these can be interpreted as coming from specific priors.

\textbf{Tikhonov.} 
Tikhonov regularization is the most commonly-used regularization both in SVMs and GPs, defined as
\begin{align}
\label{def:krr}
    \mathcal{R}(\bm{\mathcal{W}}_1,\ldots,\bm{\mathcal{W}}_D) \coloneq \, \left\lVert \bm{L} \bm{w} \right\rVert_2^2.
\end{align}
Using this kind of regularization with $\bm{L} \coloneq \lambda\,\bm{I}$ together with a kernel function is known as kernel ridge regression. From a Bayesian perspective equation~\eqref{def:krr} represents a zero-mean Gaussian prior on $\bm{w}$ with precision matrix $\bm{L}^\intercal\bm{L}$. As the total number of model parameters in a TNKM depends on the TN-ranks, overfitting becomes more likely with higher ranks. Tikhonov regularization alleviates this risk, making the final result less sensitive to ``badly chosen" TN-ranks by adding the term
\begin{align*}
    \bm{U}_d^\intercal\bm{L}^\intercal\bm{L}\,\bm{U}_d \; \textrm{vec}(\bm{\mathcal{W}}_d),
\end{align*}
to the ALS normal equations~\citep{chen2017parallelized}. The $\bm{U}_d$ matrix needs to be recomputed after each update. When $\bm{L}\coloneq \lambda\,\bm{I}$, the non-uniqueness of a TT can be exploited to turn $\bm{U}_d$ into a semi-orthogonal matrix. This reduces the regularization term to $\lambda\,\bm{I}\textrm{vec}(\bm{\mathcal{W}}_d)$ and completely avoids recomputing $\bm{U}_d$.

\textbf{Independent cores.}
Instead of putting a prior on the whole $\bm{w}$ vector, one can also consider independent priors on each of the separate TN-cores $\bm{\mathcal{W}}_d$. The Tikhonov regularization term then becomes
\begin{align*}
    \mathcal{R}(\bm{\mathcal{W}}_1,\ldots,\bm{\mathcal{W}}_D) \coloneq \sum_{d=1}^D  \left\lVert \bm{L}_d\,\textrm{vec}(\bm{\mathcal{W}}_d) \right\rVert_2^2,
\end{align*}
with $\bm{L}_d^\intercal\bm{L}_d \;\textrm{vec}(\bm{\mathcal{W}}_d)$ the corresponding term in the ALS normal equations. Choosing $\bm{L}_d\coloneq \lambda_d\,\bm{I}$ has been shown to make the final identified model insensitive to ``badly chosen" TN-ranks~\citep{MEMMEL20237294}.

A particular useful choice for $\bm{L}_d$ is the $\alpha$th-order difference matrix in the case of B-spline basis functions~\citep{karagoz2020nonlinear}. Penalizing large differences between consecutive TN-core parameters smoothens each of the univariate polynomials modeled by the B-splines. In this way, the smoothness of the predictor $\hat{y}_n$ as a whole can be ensured.

Each TN-core can also be made sparse by using an L1 regularization term~\citep{zhang2018sparse} of the form
\begin{align*}
    \mathcal{R}(\bm{\mathcal{W}}_1,\ldots,\bm{\mathcal{W}}_D) \coloneq \sum_{d=1}^D \lambda_d \, \left\lVert \textrm{vec}(\bm{\mathcal{W}}_d) \right\rVert_1,
\end{align*}
which has been shown to make the TN-cores up to 60\% more sparse without negatively impacting predictive performance.

\textbf{Automatic Relevance Determination (ARD).}
The number of univariate basis functions $I$ and TN-ranks are hyperparameters that are usually determined through cross-validation. In~\citep{kilic2025interpretable} a fully Bayesian approach was developed for CPD-based TNKMs. All parameters and hyperparameters ($I$, CP-rank $R$, variance of disturbance $e_n$, prior hyperparameters) are modeled as random variables and learned through Bayesian inference. By assuming that the posterior distribution factors over all parameters and hyperparameters, one can derive a Bayesian ALS algorithm. The key difference with the conventional ALS is that in its Bayesian counterpart one TN-core is updated while replacing all other random variables with their expectation. The Bayesian ALS has the same computational cost $O(DNIR+I^3R^3)$ as the conventional ALS using a CPD parameterization but comes with the benefit of providing uncertainty quantification. In this context a tensor-variate prior can be designed on $\bm{\mathcal{W}}_d$ that learns the dimensions of this tensor. Taking a TT parameterization as an example~\citep{afraTT}, we consider the following zero-mean Gaussian prior
\begin{align*}
    p\left(\textrm{vec}(\bm{\mathcal{W}}_d)\right) &= \mathcal{N}\left(\bm{0}, \bm{\Lambda}_{d}^{-1} \otimes \bm{\Delta}_d^{-1} \otimes \bm{\Lambda}_{d-1}^{-1} \right).
\end{align*}
Each of the precision matrices $\bm{\Lambda}_{d-1} \in \mathbb{R}^{R_{d-1} \times R_{d-1}},\bm{\Lambda}_{d}\in \mathbb{R}^{R_{d} \times R_{d}},\bm{\Delta} \in \mathbb{R}^{\times I}$ is a diagonal matrix of random variables which is updated in the Bayesian ALS. The $r_d$th diagonal entry of $\bm{\Lambda}_d$ ``controls" the $r_d$th matrix slice of the $R_d$ dimension. Should one of these precisions become large, then that particular term of $R_d$ is driven to zero, effectively truncating the rank. The same principle holds for all other TT-ranks and basis functions for each of the $\bm{x}_n$ variables. In this way TNKMs become more interpretable, as the most relevant model components are automatically identified. 

\section{Developments}
\label{sec:dev}
This section presents a brief overview of some recent TNKM developments.

\textbf{Rank-$K$ CPD feature vector.} The Kronecker product structure of the feature vector $\bm{\varphi}(\bm{x})$ is a rank-1 CPD. This can be generalized to a rank-$K$ CPD 
\begin{align}
\label{def:fl}
    \sum_{k=1}^K\, \lambda_k \; \bm{\varphi}_D^k \otimes \cdots \otimes\bm{\varphi}_2^k \otimes  \bm{\varphi}_{1}^k,
\end{align}
where each $\bm{\varphi}_d^k$ is a different feature vector. In this way different choices of basis functions can be combined into a single feature vector and their individual ``weights" $\lambda_k$ can also be estimated efficiently from the identification data using ALS~\citep{pmlr-v258-saiapin25a}. A major advantage of this approach is that cross-validation on $K$ different choices for basis functions can be completely avoided. Identifying one TNKM with a rank-$K$ feature vector as in~\eqref{def:fl} suffices, speeding up the whole identification process. In addition, a regularization term on the $\lambda_k$ parameters  effectively reduces the sensitivity of the final model to ``badly chosen" basis functions.

\textbf{Limiting Gaussian Processes.} In~\citep{pmlr-v258-wesel25a} it is shown how the 1-step ahead predictor $\hat{y}_n$ of both the CPD-based and TT-based TNMKM converge to a Gaussian Process in the limit of large TN-ranks. A crucial assumption for this convergence is that $\bm{w}_d^{r}$ for the CPD-case and $\bm{w}_d^{r_{d-1},r_d}$ for the TT-case are independent and identically distributed random variables, which is satisfied for both independent cores and ARD regularization. An important result is that for equal number of basis functions $I$, number of features $D$ and number of tensor network parameters, the TT-based 1-step predictor will converge at a polynomially faster rate than the CPD-based predictor, thus exhibiting GP behavior with a reduced number of model parameters.

\textbf{Laplace approximation.} An alternative for the fully Bayesian approach, present in the discussion on the Automatic Relevance Determination regularization, is to approximate the posterior of the TN-cores by a Gaussian distribution~\citep{saiapin2026laplace}. This also allows TNKMs to provide principled uncertainty estimates. Combining this approximation with a linearized predictive distribution has been shown to consistently outperform Gaussian Processes and Bayesian Neural Networks. 

\textbf{Recursive identification with Kalman filter.} Writing a conventional SVM model in state space form
\begin{align*}
    \bm{w}_{n+1} &= \bm{w}_{n},\;     y_n = \bm{\varphi}(\bm{x}_n)^\intercal\;\bm{w}_n +e_n,
\end{align*}
together with assuming Markovianity and a Gaussian prior $p(\bm{w}_0)$ allows one to compute the filtering density $p(\bm{w}_n | y_{1:n})$ exactly with a Kalman filter. Replacing the SVM model with a TNKM then implies that the Kalman filter update equations involve vectors and matrices of dimensions $I^D$. Fortunately, a complete Kalman filter update step can be done efficiently by computing all quantities directly as tensor networks~\citep{tnkalman}. This is yet another way to arrive at a probabilistic TNKM. More recently a square-root version of such a tensor network Kalman filter has been developed with improved numerical stability~\citep{menzen2026tensor}.

\section{Future work and Conclusion}
TNKMs have been shown to be a flexible model class and efficient in terms of storage and computation. Upcoming work focuses on the derivation and implementation of an efficient TNKM-based prior for smooth and decaying high-order Volterra kernel identification and a proper initialization to avoid unstable gradients during identification.

Some pointers for future work. The identification of different colored noise models in the TNKM context remains largely unexplored. The parameterization of a completely separate noise dynamics such as in nonlinear Box-Jenkins models has also not been tackled yet. Another largely unexplored terrain is the use of TNKMs for nonlinear state space models where the state is partially or even completely hidden. 

\begin{ack}
This publication is part of the project Sustainable learning for Artificial Intelligence from noisy large-scale data (with project number VI.Vidi.213.017) which is financed by the Dutch Research Council (NWO).
\end{ack}




\bibliography{bibliography.bib}  

\end{document}